\documentclass[10pt,preprint]{aastex}
\usepackage{graphicx}

\begin{document}
\newcommand{\bp}{$\beta$~Pic}

\title{Ro-vibrational CO Detected in the $\beta$~Pictoris Circumstellar Disk}

\author{Matthew R. Troutman\altaffilmark{1}}
\affil{
Department of Physics \& Astronomy, University of Missouri - St. Louis, 8001 Natural Bridge Rd., St. Louis, MO, 63121; mtroutm@clemson.edu
}

\author{Kenneth H. Hinkle}
\affil{
National Optical Astronomy Observatory, 950 North Cherry Avenue, Tucson, AZ 85719; khinkle@noao.edu
}

\author{Joan R. Najita}
\affil{National Optical Astronomy Observatory, 950 North Cherry Avenue, Tucson, AZ 85719; najita@noao.edu
}

\author{Terrence W. Rettig}
\affil{
Center for Astrophysics, University of Notre Dame, Notre Dame, IN 46556; trettig@nd.edu
}

\author{Sean D. Brittain}
\affil{
Department of Physics \& Astronomy, 118 Kinard Laboratory, Clemson University, Clemson, SC 29634; sbritt@clemson.edu
}

\altaffiltext{1}{Michelson Graduate Fellow}

\begin{abstract}
We present high resolution near-infrared spectra of $\beta$~Pictoris - a nearby young star with a debris disk. Fundamental low-J CO absorption lines are detected and strict upper limits are placed on the flux of v=2-1 low-J CO emission lines. The limit on the UV fluorescent emission flux in the v=2-1 lines is used to place a tight constraint on the inner extent of the CO gas. Assuming \ion{H}{1} is the primary collision partner, the subthermal population of the low-J v=0 rotational levels constrains the density of the gas in the disk to $n_{\rm H} = (2.5^{+7.1}_{-1.2})\times10^5$~cm$^{-3}$. If the distribution of hydrogen follows that of the other metals in the disk, we find that the mass of the gas in the disk is $(0.17^{+0.47}_{-0.08})$ M$_\earth$. We compare this mass to the gas mass necessary to brake the metals in the disk through ion-neutral reactions. 
\end{abstract}


\section{INTRODUCTION}

$\beta$ Pictoris ($\beta$~Pic) is a 12 Myr \citep{2001ApJ...562L..87Z}, A5V star at a distance of 19.3 pc \citep{1997A&A...320L..29C} with a debris disk. The dust component of the disk has been studied extensively from the optical through the sub-millimeter  \citep{1984Sci...226.1421S, 1989ApJ...337..494A, 1993ApJ...411L..41G, 1995AJ....110..794K, 1997MNRAS.292..896M, 1997A&A...327.1123P, 1998Natur.392..788H, 2000ApJ...539..435H, 2003A&A...402..183L, 2004Natur.431..660O, 2006AJ....131.3109G, 2009A&A...495..523B}. Asymmetries in the dust distribution, and in particular warps in the disk \citep[e.g.][]{1995AJ....110..794K,1997MNRAS.292..896M, 2000ApJ...539..435H}, point to the presence of a planet in the disk which was recently imaged \citep{2009A&A...493L..21L,2010Sci...329...57L}.

In addition to the dust, the disk has a reservoir of atomic gas with at least two components \citep[e.g.][]{1987A&A...185..267F,1998A&A...329.1028J,2000ApJ...538..904R,2004A&A...413..681B}. One component shows lines red-shifted relative to the rest frame of $\beta$ Pic, resulting from infalling gas released from star-grazing planetesimals \citep[e.g.,][]{1994A&A...290..245V}. A second component (comprised of \ion{Na}{1}, \ion{Fe}{1}, \ion{Ca}{2}, \ion{Ti}{2}, \ion{Ni}{1}, \ion{Ni}{2}, \ion{Cr}{2})  remains at rest relative to the star and has a spatial distribution similar to that of the dust  \citep{1987A&A...185..267F,2001ApJ...563L..77O,2004A&A...413..681B}. The existence of the stable gas in a debris disk was somewhat surprising because radiation pressure from $\beta$~Pic to should be sufficient to radially accelerate the gas to high velocities \citep[$\gtrsim100$~km~s$^{-1}$;][]{2004A&A...413..681B}. Thus, some braking mechanism is necessary to explain the stable component of the gas. \citet{2004A&A...413..681B} estimated that $\sim 50 M_\earth$ of gaseous hydrogen is needed if it is the braking agent. However, \citet{1995A&A...301..231F} show from limits on 21 cm emission that the atomic hydrogen mass must be less than 1.6 M$_\earth$. As for the molecular gas, \cite{2001Natur.412..706L} place a tight constraint on the mass of H$_2$ at $<$0.1 M$_\earth$ based on the non-detection of absorption lines from electronic transitions. Additionally, \citet{2007ApJ...666..466C} use the absence of H$_2$ emission in Spitzer IRS spectra to limit the amount of warm H$_2$ (50 - 100 K) in the disk at $<$17 M$_\earth$, ruling out a more extended reservoir of molecular gas. \citet{2005A&A...437..141T} use dynamical arguments to argue that the total gas mass must be $<$0.4 M$_\earth$. Although these observational constraints would appear to suggest that the disk is not massive enough to brake the gas, \citet{2006ApJ...643..509F} note that a much smaller mass of hydrogen may be sufficient to brake the gas because only the ionized component needs to be decelerated. This is because the elements that receive the highest radiative force are also the elements with the highest ionization rates, thus only the ions are important. The acceleration of the neutrals is negligible, thus they do not need braking. \citet{2006ApJ...643..509F} explored possibilities for braking the ions in the gas, which would explain the observed stable component of the gas.

The first possibility is ionized particle collisions, where the ions lose momentum by Coulomb interactions. The ions are dynamically coupled, so the radiative force on the particles can be described by an average over all of the particles. Some ions feel a low radiative force, so if these atoms were somehow overabundant, the average radiative force on the ions would go down - effectively slowing the gas. Carbon is one such atom that feels a low radiative force. Indeed, an increase in the carbon abundance by a factor of $\sim$10 with respect to other atoms may be sufficient to lower the average radiative force on the ions, braking the gas. Atomic carbon is observed to be enhanced compared to oxygen in the $\beta$~Pic system by a factor of 18 relative to solar \citep{2006Natur.441..724R}, making this an attractive possibility.

Another possibility for braking the gas is ion collisions with neutral gas. In this scenario, a total disk mass of at least 0.1 M$_\earth$ is required \citep{2006ApJ...643..509F}, well within all of the observational limits mentioned above. This is much smaller than the value inferred by \citet{2004A&A...413..681B}, as they did not assume only the ionized particles need to be slowed. While the ion-ion collision scenario is possible, limits on the disk mass cannot rule out a contribution to the stabilization of the atomic gas from ion-neutral collisions as well.

To explore the role ion-neutral collisions play in braking the gas in the outer disk around $\beta$~Pic, we present near infrared high resolution spectra of $\beta$~Pic centered near 4.7 $\mu$m. We detect fundamental ro-vibrational CO lines in absorption and place strict constraints on fluorescent emission from higher vibrational bands. We find that the rotational levels are only thermalized up to J$''$=2 and describe how this information provides a means to calculate the density of gas in the disk. We also use the non-detection of fluoresced ro-vibrational CO emission to constrain the inward extent of CO in the disk.

\section{OBSERVATIONS}\label{sec:obs}
We acquired high-resolution near-infrared spectra of $\beta$~Pic using PHOENIX \citep{2003SPIE.4834..353H,2000SPIE.4008..720H,1998SPIE.3354..810H} at the Gemini South telescope and CSHELL \citep{1990SPIE.1235..131T} at the NASA Infrared Telescope Facility. The resolutions of PHOENIX and CSHELL are R=50,000 and R$\sim$40,000, respectively. The PHOENIX observations were taken on March 23, 2008 and January 12, 2003. The CSHELL observations were taken on August 8, 2000. The observations were centered around 4.7~$\mu$m to cover fundamental ro-vibrational transitions of CO. A summary of observations, presented in Table \ref{tab:observations}, includes the lines that are observed. 

\begin{figure}
\plotone{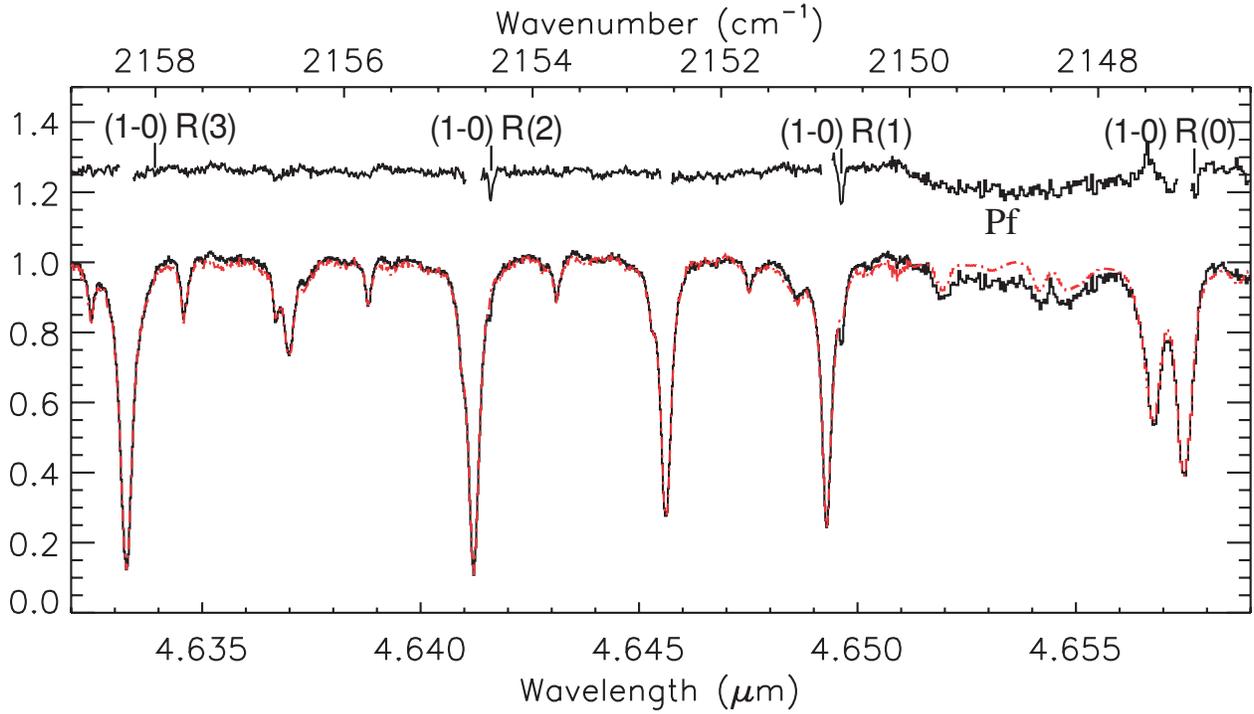}
\caption{The spectrum of $\beta$ Pic, including both the PHOENIX ($<4.651\mu$m) and CHSHELL ($>4.651\mu$m) observations. The spectrum $\beta$ Pic (solid line) and the telluric standard (dot-dashed line) are plotted in normalized units. The ratio of the spectra is also plotted and offset by 0.25 units. Areas where the transmittance is less than 50\% are omitted. The locations of the CO transitions are shown for v=1-0. Absorption is observed from the low-J, v=1-0 levels. }
	\label{fig:BetaPicSpectrum}
\end{figure}

\begin{figure}
\plotone{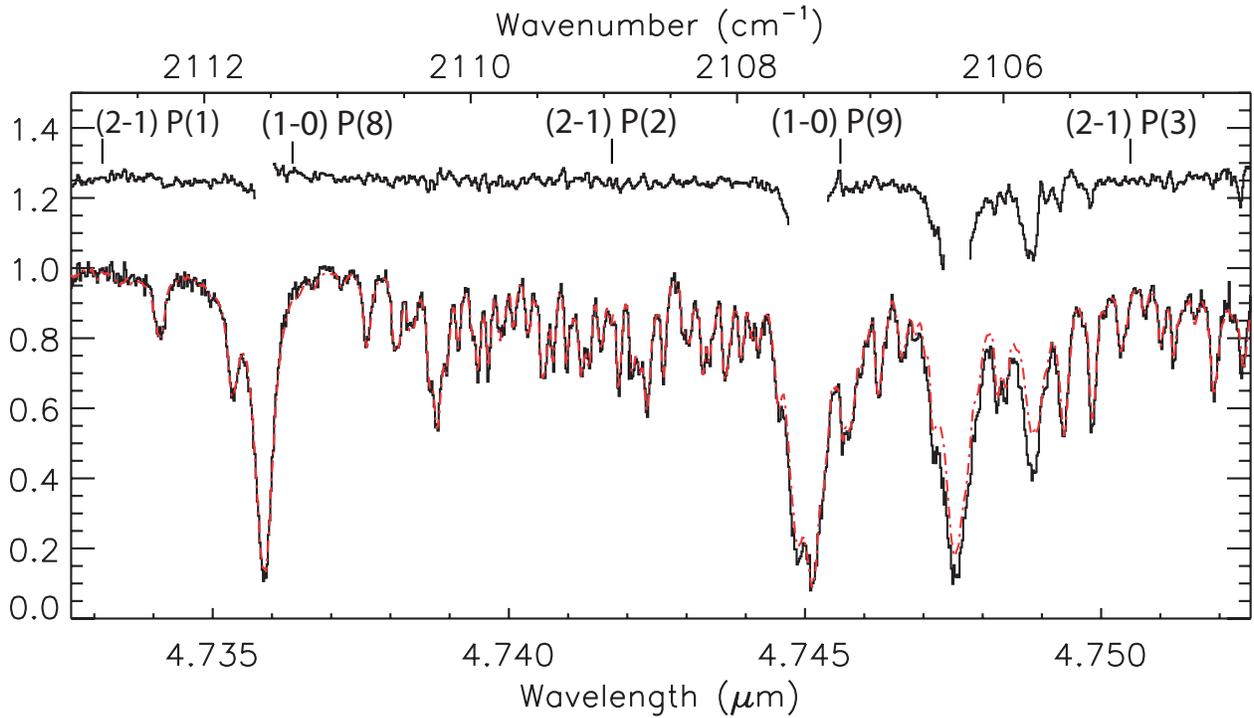}
\caption{The spectrum of $\beta$ Pic. The spectrum $\beta$ Pic (solid line) and the telluric standard, HR~5671 (dot-dashed line), have been normalized and overplotted. The ratio of the spectra is also plotted and offset by 0.25 units. Areas where the transmittance is less than 50\% are omitted. The locations of the CO transitions are shown for v=1-0 and v=2-1. No emission is observed, including the v=2-1 P(1) line at $\sim4.733$ $\mu$m. There appears to be a bump at the location of the v=1-0 P(9) feature, yet this is embedded in a combination of telluric features. Additionally, if the P(9) feature was a detection, the P(8) feature would also show emission, which is not observed. }
	\label{fig:BetaPic2}
\end{figure}

Observations in the 4.7~$\mu$m region are dominated by a strong thermal background, limiting the exposure times. Therefore, short  exposures  are taken while nodding between two positions separated by $\sim$5$\arcsec$. The exposures are taken in an ABBA pattern in order to cancel the thermal continuum to first order. The scans are flat fielded, cleaned of hot and dead pixels, including cosmic ray hits, and then combined in the sequence (A$_1$-B$_1$-B$_2$+A$_2$)/2. Because the spectra are curved along the detector, they are first rectified by finding the centroid of each column and shifted to a common row. A one-dimensional spectrum is then extracted from the rectified spectrum. This spectrum is combined with an atmospheric transmittance model spectrum in order to find a wavelength solution. The model is created using the Spectral Synthesis Program \citep{1974JQSRT..14..803K}, which accesses the 2000HITRAN molecular database \citep{2003JQSRT..82....5R}. Each spectrum is then ratioed to a standard star observed at a similar airmass to remove telluric absorption lines. Areas where the transmittance is below 50\% are omitted. To determine the signal to noise of our spectra, we measured the standard deviation of continuum (Table \ref{tab:observations}). The extracted spectra are presented in Figures \ref{fig:BetaPicSpectrum} and \ref{fig:BetaPic2}. The signal to noise of the spectrum acquired with PHOENIX was 100. The spectrum acquired with CSHELL is dominated by the broad Pf $\beta$ feature and mis-canceled water line. The signal to noise of the spectrum acquired with CSHELL was 90.

\section{RESULTS} \label{sec:results}
We detect the R(0), R(1), and R(2) fundamental ro-vibrational CO absorption lines near 4.64 $\mu$m (Figure \ref{fig:BetaPicSpectrum}). We do not detect the R(3) feature. The R(1) and R(2) lines are unresolved at the the 6 km s$^{-1}$ resolution of PHOENIX, which is consistent with the intrinsic line width b=1.3 km s$^{-1}$  inferred from the electronic absorption measurements observed in the UV \citep{2000ApJ...538..904R}. Similarly, the R(0) line observed with CSHELL was unresolved at the effective $\sim$8 km s$^{-1}$ resolution of the instrument. The R(1) was not detected with CSHELL as the CO line was not cleanly separated from the heavily saturated telluric component. The N$_2$O line on the red wing of the telluric feature lowers the transmittance of the atmosphere an additional 20\%.  The heliocentric Doppler shift of the CO absorption lines for all dates are $+21\pm1$ km s$^{-1}$, consistent with the heliocentric Doppler shift of $\beta$~Pic, and indicates that it is part of the stable component of the gas disk \citep{1987A&A...185..267F,1998A&A...329.1028J,2000ApJ...538..904R,2004A&A...413..681B,2006Natur.441..724R,2008ApJ...676..509R}. 

Figure \ref{fig:BetaPic2} shows the spectrum in the 4.74 $\mu$m region containing the v=1-0 P(8) and P(9) lines and the v=2-1 P(1), P(2), and P(3) lines. These features are not detected. While the gas in the disk around $\beta$~Pic is too cold to give rise to collisionally excited CO ro-vibrational emission or absorption, CO ro-vibrational emission can be excited by UV fluorescence \citep{1980ApJ...240..940K}. When CO is excited electronically and relaxes back to the ground electronic state, the excited vibrational bands are populated. Thus, even relatively cool gas can give rise to a large vibrational temperature. For example, the CO around HD~141569 has a rotational temperature of 200~K while the vibrational temperature is $\sim$5000~K \citep{2003ApJ...588..535B,2007ApJ...659..685B}. CO was first detected around $\beta$~Pic in the UV via electronic absorption lines. Thus, one may expect to observe cold ro-vibrational emission from $\beta$~Pic. We do not detect this emission in our spectrum around 4.74 $\mu$m, with a signal-to-noise of 90. There appears to be a spurious peak at the at the location of the v=1-0 P(9) feature. However, this is not assumed to be a true featue. The peak is embedded in a combination of telluric features, with the transmittance at the location only at 68\%. Additionally, the P(9) emission cannot be significantly brighter than the non-detected P(8) feature. Finally, the gas would have to be quite warm to detect it this far up the rotational ladder, which is also not observed (see Section \ref{sec:ro_vib_abs}). Assuming the lines are Gaussian and unresolved, our one-sigma upper limit on the equivalent width (based on the S/N of the spectrum) is $\sim5\times10^{-4}$ cm$^{-1}$. Scaling the equivalent width by the flux density as measured by the Infrared Space Observatory (F$_{4.7 \mu m}=1\times10^{-9}$ ergs s$^{-1}$ cm$^{-2}$ $\micron^{-1}$), the upper limit on the flux of the individual lines is $7.1\times10^{-16}$ ergs cm$^{-2}$ s$^{-1}$ (Table \ref{tab:fluxes}).

\section{ANALYSIS}\label{sec:analysis}
\subsection{Constraint on CO Distribution}\label{constraint}
The excitation of the electronic transitions of CO via the absorption of UV flux results in the population of excited vibrational levels in the ground electronic state \citep[e.g.][]{1980ApJ...240..940K}. For example, \citet{2003ApJ...588..535B} observe ro-vibrational CO emission from the disk around HD~141569 distributed from the inner 50~AU and conclude that the vibrational temperature of the gas is $\sim$5500~K while the rotational temperature of the gas is $\sim$200 K. This situation occurs when the vibrational population of cool gas is dominated by the de-excitation of electronically excited molecules. In this ``strong pumping limit,'' the vibrational population of CO near the star reflects the color temperature of the star's ultraviolet field \citep{1980ApJ...240..940K}. Further from the star, the spontaneous de-excitation of the vibrational levels becomes more significant relative to the rate at which these levels are populated by the relaxation of electronically excited molecules and the vibrational temperature decreases.

We can use the non-detection of the CO emission in the disk to constrain the distribution of CO using the upper limits on the v=2-1 emission features (displayed in Figure \ref{fig:BetaPic2}) and the analysis in the previous section. The fluxes of the v=2-1 P(1), P(2) and P(3) lines are $<7.1\times10^{-16}$ erg~cm$^{-2}$~s$^{-1}$ (Section \ref{sec:results}).

\citet{1998A&A...329.1028J} and \citet{2000ApJ...538..904R} observe the CO absorption lines near 1500\AA\ that represent transitions from the ground electronic state to the first excited electronic state. When these molecules relax, they will populate v$\geq$1 in the ground electronic state. We have calculated the population of v=2 for CO as a function of distance from the star following the procedure described by \citet{2009ApJ...702...85B}. We take our input UV spectrum for $\beta$~Pic taken from Kurucz atmospheric models \citep{1993ASPC...44...87K}, which are then scaled to the emission presented in \citet{1998A&A...329.1028J}. 

As a limiting case, we first assume that the CO in the disk is distributed over the same radial range as the atomic gas observed by \citep{2004A&A...413..681B}. The UV is absorbed over an annulus extending from 13 AU to 323 AU, with a profile determined by \citet{2004A&A...413..681B}. They infer the radial profile for the gas disk based on their measurement of the distribution of \ion{Na}{1} (discussed in detail in Section \ref{sec:excitation}):
\begin{equation}\label{eq:line_of_sight}
	n_{CO}(r)=n_0 \left[\left(\frac{r}{117{\rm \;AU}}\right)^{2.4}+\left(\frac{r}{117{\rm \;AU}}\right)^{5.3}\right]^{-1/2}
\end{equation}
where n$_0$ is the fiducial density of CO at 117 AU. 

\begin{figure}
\plotone{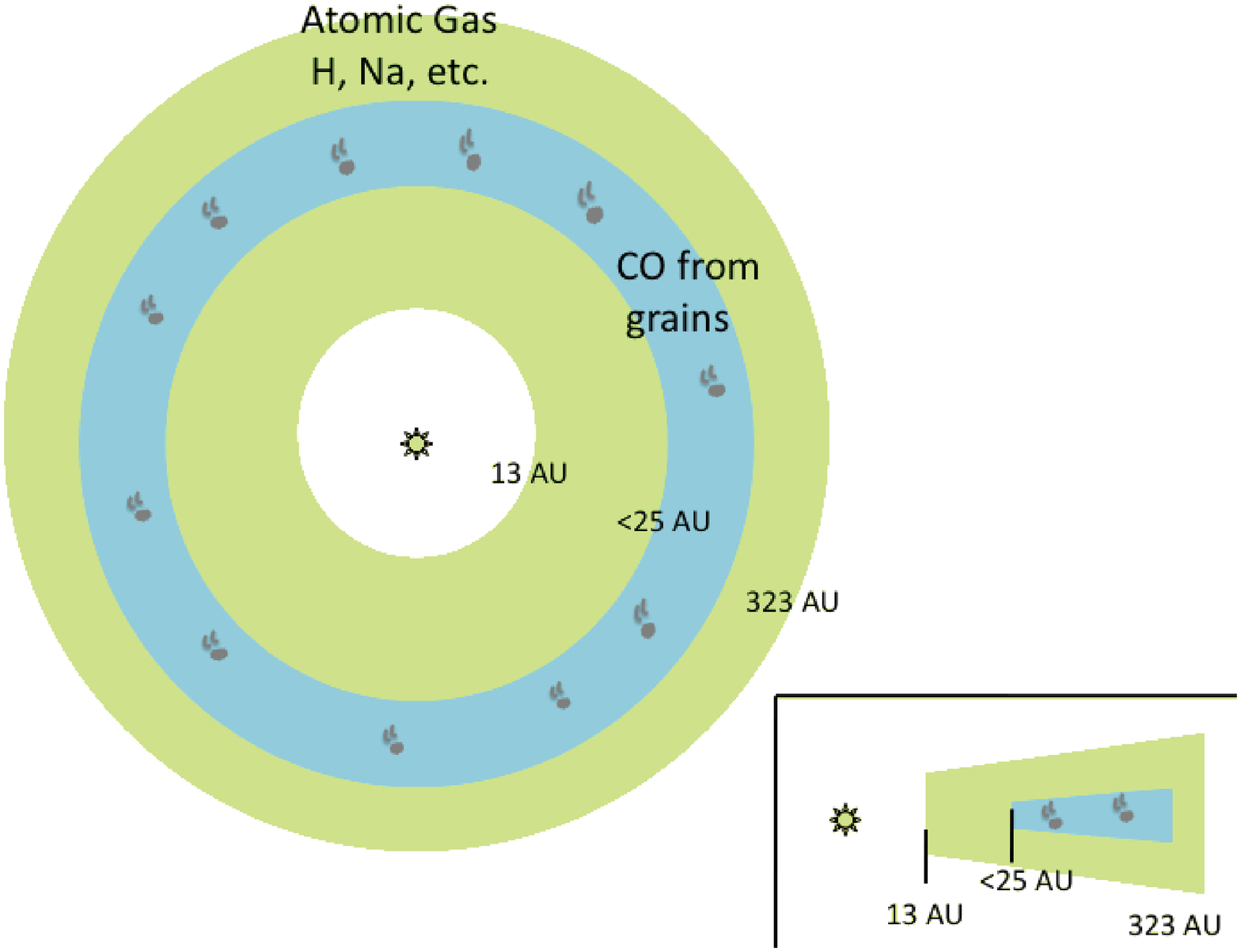}
\caption{Distribution of CO relative to other gaseous components in the $\beta$ Pic system. The atomic component of the disk (\ion{H}{1} and other metals) is observed to extend from 13 AU to 323 AU \citep{2004A&A...413..681B,2006ApJ...643..509F}. The properties of the fundamental CO absorption reported here are consistent with an origin for this molecular component as CO photodesorbed from icy grains (gray objects) with an inner radius of 25 AU and an unknown outer radius. The CO is excited by collisions with the HI component of the disk. }
\label{fig:diagram}
\end{figure}

The UV flux propagates through each gas annulus, where the flux is attenuated by the distance from the host star as it travels through the annulus. Throughout the annulus, the population of CO molecules in v=2 falls of with the distance to the star. In this case, n$_0$ is normalized to reproduce a line-of-sight column density of $2.1\times10^{15}$~ cm$^{-2}$ (calculation of the column density is described in Section \ref{sec:ro_vib_abs}). We require an inner radius r$_{\rm in}\ge25$ AU for the disk to intercept a small enough fraction of the stellar flux that the emergent UV fluorescent flux is below the observational limits. If the CO is restricted to a smaller radial extent, the number of molecules in v=2 will increase as more gas is nearer the star. In this case, the inner edge of the gas is at a larger radius. Thus, the inner radius must be larger than 25 AU. The assumption that the CO has the same radial distribution as the \ion{Na}{1} is a limiting case. If a different profile is used, for example something closer to a Kuiper Belt distribution where the density of CO would fall off faster than the profile used, then the CO population in v=2 would increase closer to the star, which would increase the limit on the inner radius of CO emission to remain consistent with the non-detection of the v=2 features. Thus, the inner radius has a hard constraint at r$_{\rm in}\ge25$ AU. There is no constraint on the outer radius (Figure \ref{fig:diagram}).

\subsection{Ro-vibrational Absorption}\label{sec:ro_vib_abs}
\begin{figure}
\plotone{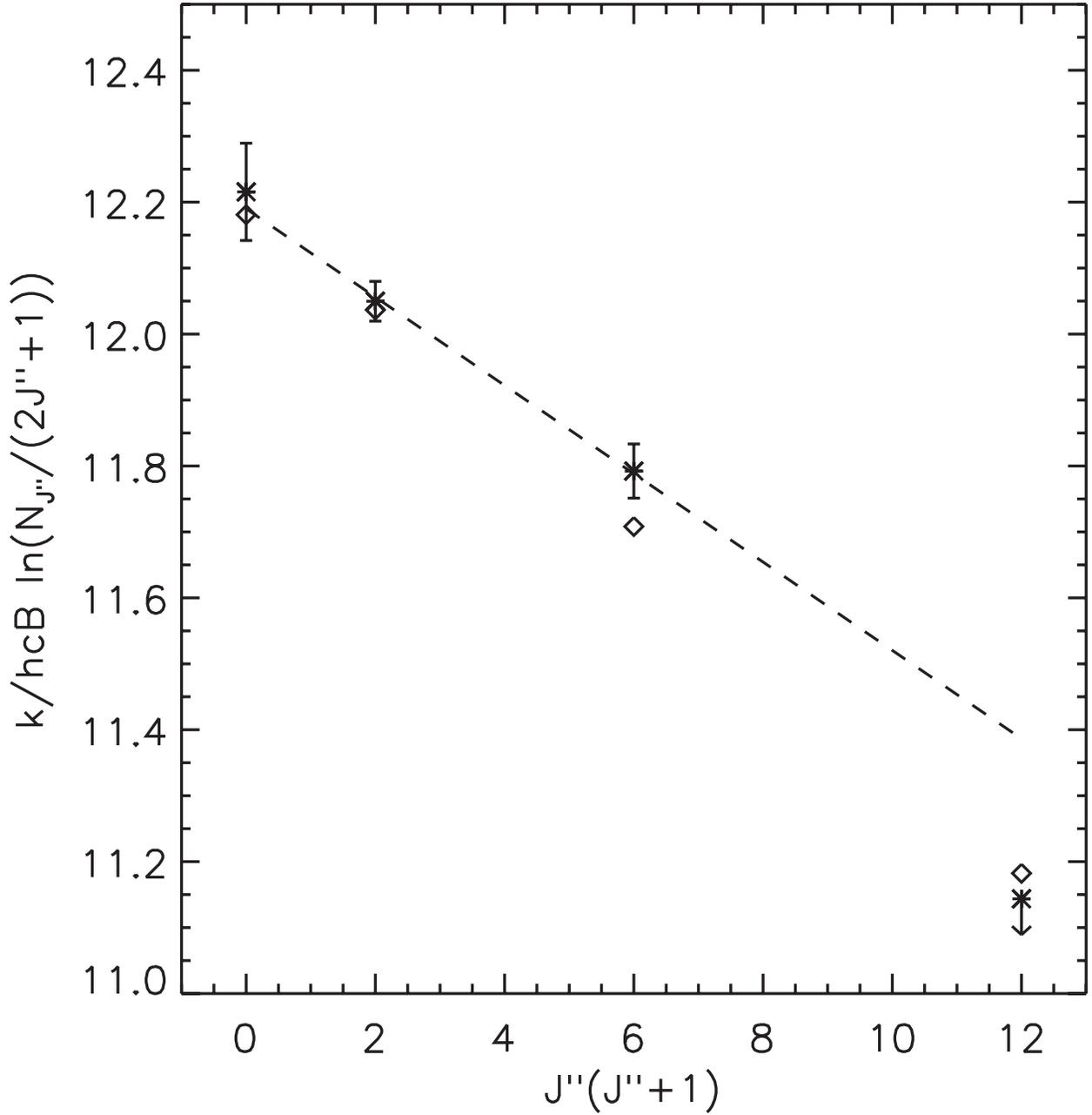}
\caption{CO excitation plot of the low-J ro-vibrational absorption lines. The data are shown as asterisks with error bars calculated from the 1$\sigma$ variation of the continuum around the features, while the best-fitting model is overplotted as diamonds. The dashed line represents a temperature of $15\pm2$ K. The R(3) feature is an upper limit, and much lower than the LTE fit to the lower J lines, suggesting that J$''=3$ is sub-thermally populated. The sub-thermal population allows the calculation of the density of the collision partner. }
\label{fig:thermalj}
\end{figure}

We can also use the equivalent width of the absorption features to determine the rotational temperature of the absorbing gas. The column density of molecules in each state is related to the equivalent width of absorption by:
\begin{equation}
	N_{J''}=\frac{W_{\tilde{\nu}}}{8.85\times 10^{-13} f_{J'J''}}\end{equation}
where J$''$ is the lower transition state, $W_{\tilde{\nu}}$ is the equivalent width of the line, $8.85\times10^{-13}$ cm is the classical electron radius ($\pi$e$^{2}$/m$_e$c$^{2}$), and $f_{J'J''}$ is the oscillator strength of the transition. The equivalent widths and column densities are presented in Table \ref{tab:fluxes}. If the CO is in local thermodynamic equilibrium, then the relative population of each level is given by the Boltzmann distribution, 
\begin{equation}
	N_{J''}=\frac{N_{\rm tot}(2J''+1)}{Q}e^{-hcBJ''(J''+1)/kT},
\end{equation}
where $Q$ is the rotational partition function, and $B$ is the rotational constant. This can be rewritten as:
\begin{equation}
	\frac{k}{hcB}ln\left(\frac{N_{J''}}{2J''+1}\right)=-\frac{1}{T}J''(J''+1)+\frac{k}{hcB}ln\left(\frac{N_{\rm tot}}{Q}\right),
\end{equation}
where $N_{\rm tot}$ is the total number of CO absorbing molecules. Thus, plotting $(k/hcB)\;ln(N_{J''}$/2J$''$+1) vs J$''$(J$''$+1) results in a linear relationship where the negative reciprocal of the slope is equal to the rotational temperature. However, when we plot our data (Figure \ref{fig:thermalj}), we find that while the first three levels are consistent with a rotational temperature of $15\pm2$ K, the column density of molecules in J$''=3$ is under-populated by a factor of 2.6. A sub-thermal population can occur when the gas density is too low for collisions to balance spontaneous emission. The near-infrared transitions are optically thin, so we sum the column densities from J$''=0$ to J$''=2$ to find the column density of CO, $(2.1\pm0.3)\times10^{15}$~cm$^{-2}$, with the column density of CO in J$''=3$ is $<2.3\times10^{14}$~cm$^{-2}$. We assume the overall CO column density to be $(2.1\pm0.3)\times10^{15}$~cm$^{-2}$. 

\citet{1998A&A...329.1028J} used electronic absorption spectra of CO from the Goddard High-Resolution Spectrograph ($\sim$1500\AA) to measure the column density and temperature of CO and found N(CO$)=(2\pm1)\times10^{15}$ cm$^{-2}$ and T=$20\pm5$~K. \citet{2000ApJ...538..904R} used the Space Telescope Imaging Spectrograph (STIS) to find a CO column density of N(CO$)=(6.3\pm0.3)\times10^{14}$~cm$^{-2}$ at a temperature of $15.8\pm0.6$ K. Our column density is consistent with that of \citet{1998A&A...329.1028J}, and a factor of $\sim$3 larger  than that reported by \citet{2000ApJ...538..904R}, while the temperature we derive is consistent with both studies. As both \citet{1998A&A...329.1028J} and \citet{2000ApJ...538..904R} are measuring absorption of the same electronic transitions, these two UV studies should be tracing the same column of gas. Further, the ro-vibrational absorption we observe should probe the column of gas observed against the continuum of the star as there is no extended Ly$\alpha$ emission in the UV \citep{1997A&A...317..521M}, so it is unclear why the column density observed on these three epochs is different, though the discrepancy between \citet{1998A&A...329.1028J} and \citet{2000ApJ...538..904R} is only 1.5$\sigma$ due to the large error on the measurement by \citet{1998A&A...329.1028J}. 

One possibility is that the absorption is variable on the 3 year timescale between the UV observations on November 1994 \citep{1998A&A...329.1028J}, and December 1997 \citep{2000ApJ...538..904R}. We do not see this degree of variability between the August 2000 CSHELL and the January 2003 Phoenix observations. If the column of CO changed by a factor of three during this time, we would not find a linear rotational temperature diagram. Indeed, a change in the column density by a factor of three between the 2000 and 2003 observations would cause the J$''=0$ point to be offset by $\sim0.4$ units relative to the other data points in our excitation diagram (Figure \ref{fig:thermalj}). Another explanation may be that the UV observations of the J$''<3$ lines are saturated,  providing an underestimate to the total column density. \citet{2000ApJ...538..904R} assumed the rotational levels were in LTE; however, our results indicate that this is not the case (see Figure \ref{fig:thermalj}). If the lowest three lines were slightly saturated ($\tau _0 \gtrsim 1$), then it is possible that the total column of CO could be significantly underestimated by assuming a thermal population while fitting to optically thin, sub-thermally populated lines. To determine if the assumption of NLTE can reproduce the observed equivalent widths of the CO fundamental lines, we calculate the rotational levels explicitly in the next section. 

\subsection{Excitation}\label{sec:excitation}
The population of the rotational levels of CO are calculated explicitly under the assumption that the molecules are excited by a combination of collisional excitation and radiative pumping by the cosmic microwave background, and that the level populations are in steady state. The lifetime of CO in the disk is $\sim$200 years \citep{2000ApJ...538..904R} based on the timescale of dissociation from interstellar UV photons, so we assume that the creation and destruction of CO in the disk is unimportant. To highlight the role these various processes play in determining the populations of each rotational level, we consider the state J=1, 
\begin{eqnarray}\label{eq:2}
	\frac{dn_{J=1}}{dt} &=& +B_{01}{\cal J}_{01} n_0-(A_{10}+B_{10}{\cal J}_{10})n_1 \nonumber \\
		&& -\sum_{i=0,i\ne1}^{7}k_{1i}n_1+(A_{21}+B_{21}{\cal J}_{21})n_2 \nonumber \\
		&& +\sum_{m=0,m\ne1}^{6}k_{m1}n_m =0,
\end{eqnarray}
where B$_{ij}$ is the Einstein coefficient, ${\cal J}_{ij}$ is the energy density of the cosmic microwave background at the frequency of the transition, and k$_{ij}$ is the collisional (de)excitation coefficient from the $i$th state to the $j$th state, and $n_i$ is the fractional population of the $i^{th}$ level. The collisional coefficients are given by
\begin{equation}
	k_{ij}=\sum n_s f_s(T), 
	\end{equation}
 for an upward transition ($i<j$), and
	\begin{equation}
	k_{ji}=\frac{2J'+1}{2J''+1}e^{-hc\tilde{\nu}/kT}\sum n_s f_s(T) 
\end{equation}
for a downward transition ($i>j$).

The variable $n_s$ is the number density of the collision partner, J$''$ is the lower state, J$'$ is the upper state, and $f_s(T)$ is the collision rate for a specific species, $s$. A suite of synthetic spectra were created with the number density of the collision partner as a free parameter. We assume an intrinsic line broadening of 1.3 km s$^{-1}$ \citep{2000ApJ...538..904R}. The equivalent widths of the features from the synthetic spectrum are compared to the observed equivalent widths. We determine the best fit to the density by calculating chi-squared for each density of a given collision partner. 

The CO in the disk is likely being photodesorbed from icy grains. This is a process by which the molecules in the ice lattice are electronically excited. This energy breaks the bonds of the lattice allowing the molecules to escape. The photodesorption efficiency of CO is similar to the photodesorption efficiency of H$_2$O \citep{2009ApJ...693.1209O,2009A&A...496..281O}. The mixing ratio for H$_2$O/CO can range from as low as theoretical values of 1 \citep{1985A&A...152..130D} to as high as $\sim$50 as observed in Oort cloud comets \citep{2008SSRv..138..127D}. Thus it is plausible that H$_2$O could be the dominant collision partner with CO. 

\begin{figure}
\plotone{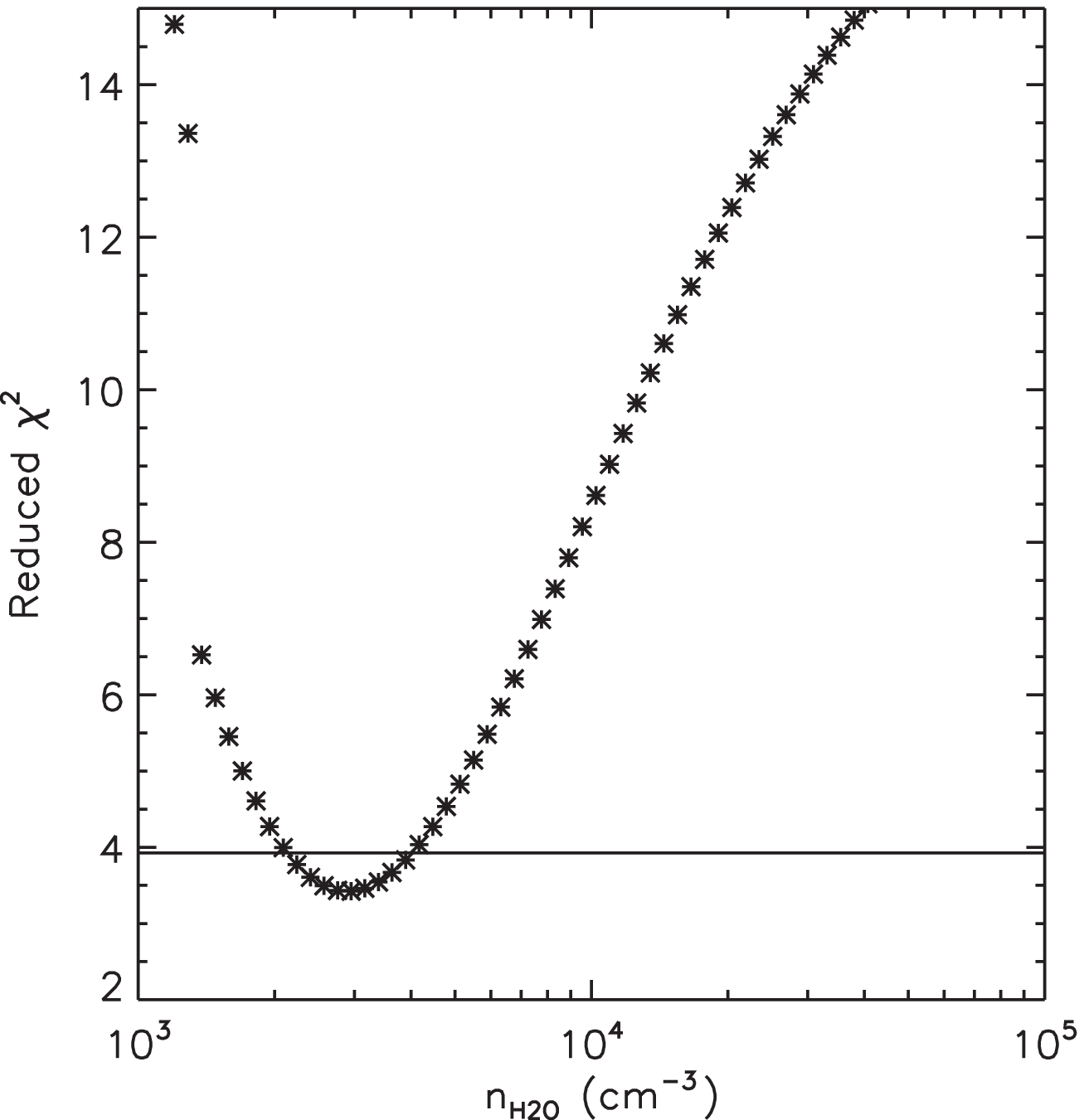}
\caption{A one-dimensional chi-squared plot for the fitting of the equivalent widths using H$_2$O as the collision partner. For each value of n$\rm_{H_2O}$, we calculated the rotational population of CO and the resultant equivalent widths of the v=0-1 R(0), R(1), R(2), and R(3) lines. The reduced chi-square statistic for each run of our model is plotted, and the 68\% confidence level corresponding to our 1$\sigma$ uncertainty is plotted as a solid line. We find that n$\rm_{H_2O}=(3.0^{+1.5}_{-1.0})\times10^{3}$ cm$^{-3}$ provides the best fit to the data.}
\label{fig:chisq_h2o}
\end{figure}

The collision rate between H$_2$O and CO has been estimated by \citet{1993ApJ...412..436G} to be $1.04\times10^{-11}$~cm$^3$~s$^{-1}$ at 15~K for the transition between the ground and first excited rotational level. Adopting these parameters, we find that the best fitting H$_2$O density is n$_{\rm H_2O}=(3.0^{+1.5}_{-1.0})\times10^{3}$~cm$^{-3}$. Figure \ref{fig:chisq_h2o} shows the chi-squared values assuming 15 K for different densities of H$_2$O. If we assume that the column density of H$_2$O is 1-50 times the column density of CO, and the CO and H$_2$O are co-spatial in an annulus, then the path length of the water annulus would be:
\begin{equation}
	L_{\rm annulus}=\frac{N(H_2O)}{n_{H_2O}}
\end{equation}	
This leads to a radial path length of the gas annulus ranging from $4.7\times10^{-2}$ AU to a maximum of 2.3 AU. The desorption efficiency depends on the intensity of the incident UV, which decreases as r$^{-2}$ in the radial direction. It seems unlikely that the desorption would decrease sharply at a distance of 25 AU from the star, restricting the gas to such a small annulus. Both CO and H$_2$O should be coming off of grains just as prodigiously at, for example, 35 AU, as at the inner radius of 25 AU. While it is true that if the disk is extended, it is implausible that H$_2$O is the dominant collision partner, observations show evidence for disks that have sculpted, narrow annuli \citep[such as Fomalhaut;][]{2008Sci...322.1345K, 2009ApJ...693..734C}. Other species, such as CH$_4$, may also photodesorb from grains and collisionally excite CO, as the photodesorption efficiencies are similar to CO and H$_2$O \citep{2009ApJ...693.1209O,2009A&A...496..281O}. Collision rates between these species and CO have not been measured, though if we assume that collision rates for species like CH$_4$ behave similarly to molecules such as H$_2$O and N$_2$ (the rates for H$_2$O were calculated from N$_2$ collisions), the same conclusion can be drawn about other molecules being a dominant collision partner. Thus, we cannot rule out the possibility that H$_2$O (or other similar molecules) can explain the excitation of CO.

An alternative scenario is that at an age of of 12 Myr, $\beta$ Pic retains some primordial hydrogen. Thus the CO could be photodesorbed from icy grains and collisionally excited by primordial hydrogen gas in the disk.  While \ion{H}{1} has not been detected, the upper limits do not rule out the possibility that the hydrogen abundance reflects a solar composition \citep{2004A&A...413..681B}. Indeed, \citet{2004A&A...413..681B} use a photoionization code to infer the radial profile of hydrogen in the gas disk based on their measurement of the distribution of \ion{Na}{1}: 
\begin{eqnarray}\label{eq:distribution}
	n_{H}(r,z)&=&n_0 \left[\left(\frac{r}{117{\rm \;AU}}\right)^{2.4}+\left(\frac{r}{117{\rm \;AU}}\right)^{5.3}\right]^{-1/2} \nonumber\\
	 		&&\times\;{\rm exp}\left(-\frac{z^2}{(0.17r)^2}\right),
\end{eqnarray}
where the metals are observed to extend from 13 AU to 323 AU. Assuming the disk reflects a solar composition, they find the mass of the disk is $\sim$0.1 M$_\earth$.This is below the detection limits determined from 21 cm observations of the disk \citep{1995A&A...301..231F}. 

\begin{figure}
\plotone{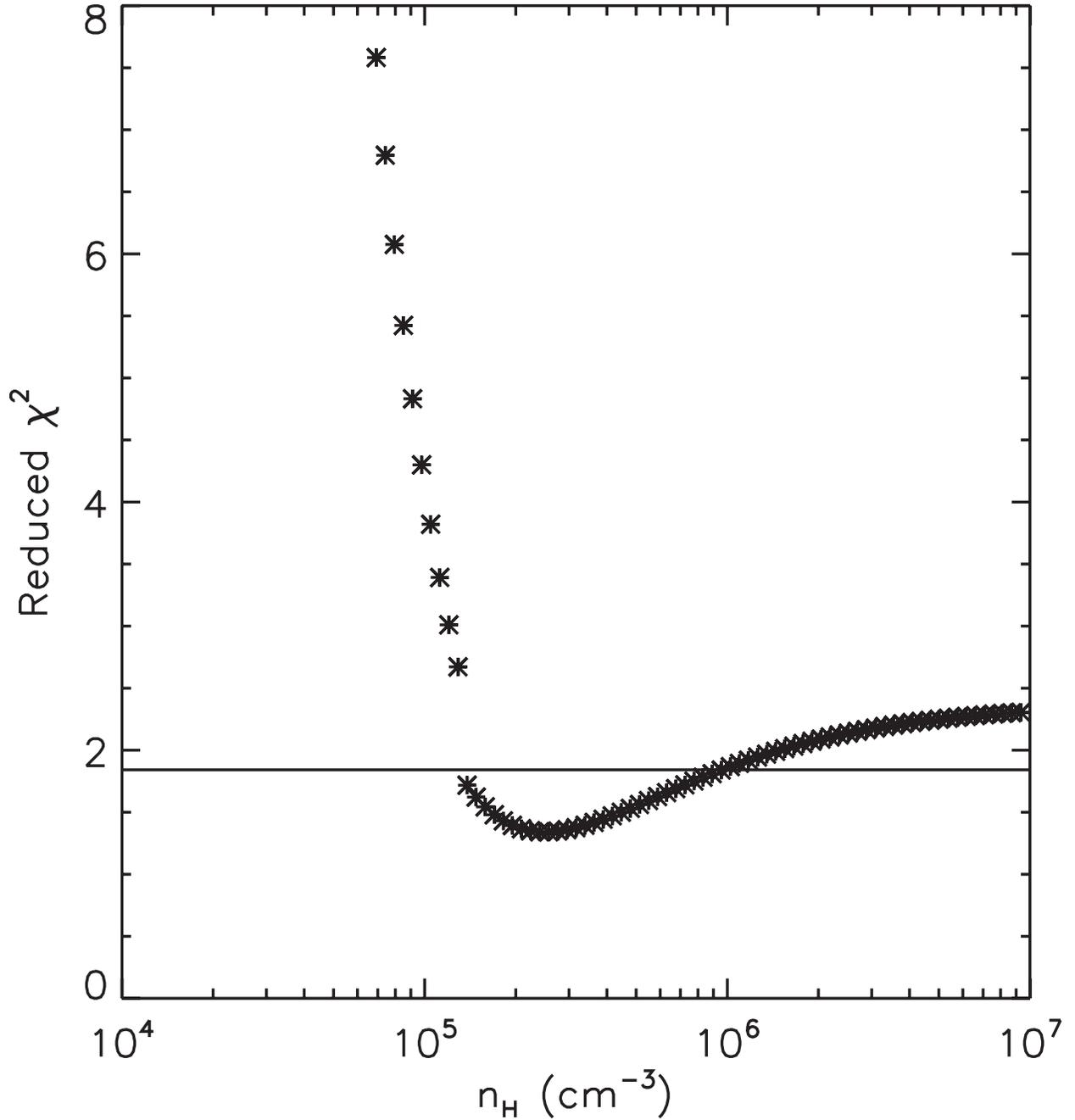}
\caption{A one-dimensional chi-squared plot for the fitting of the equivalent widths using hydrogen as the collision partner. For each value of n$\rm_H$, we calculated the rotational population of CO and the resultant equivalent widths of the v=0-1 R(0), R(1), R(2), and R(3) lines. The reduced chi-square statistic for each run of our model is plotted, and the 68\% confidence level corresponding to our 1$\sigma$ uncertainty is plotted as a solid line. We find that n$\rm_H=(2.5^{+7.1}_{-1.2})\times 10^5$ cm$^{-3}$ provides the best fit to the data. }
\label{fig:chisq_h}
\end{figure}

\begin{figure}
\plotone{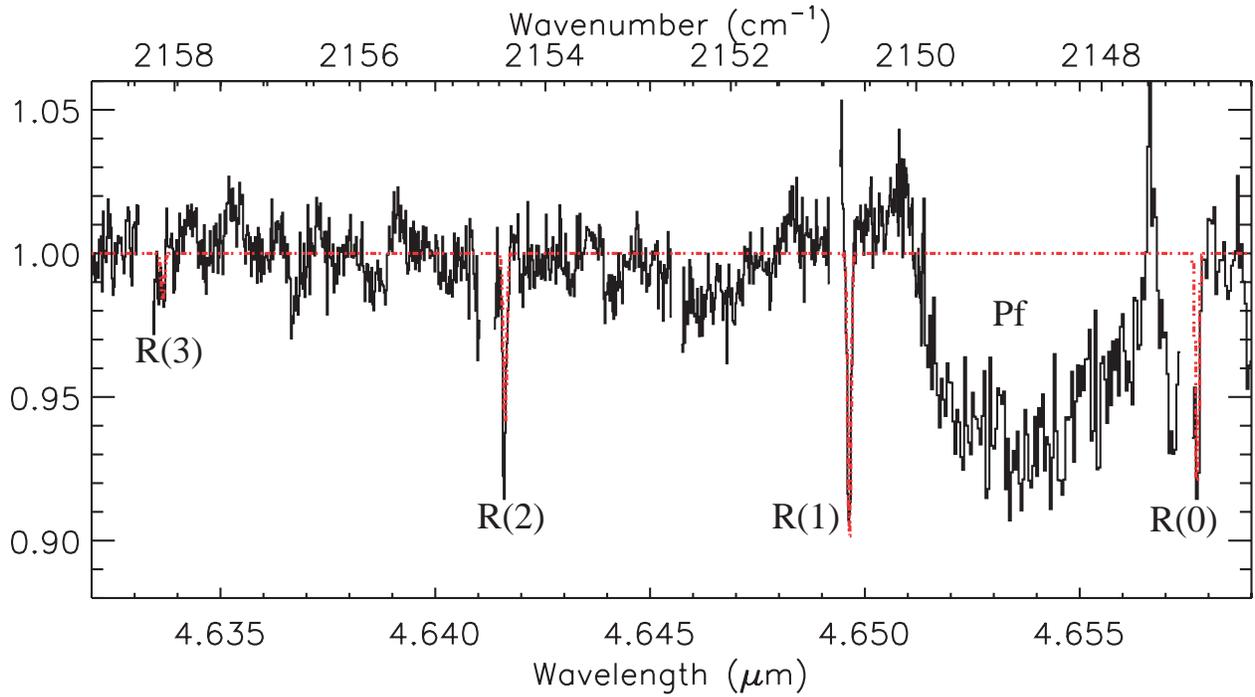}
\caption{Spectrum of CHSELL ($>4.651$ $\mu$m) and PHOENIX ($<4.651$ $\mu$m). The concatenation of the data from Phoenix and CSHELL reveals the R(0), R(1), and R(2) ro-vibrational CO absorption lines.The broad feature centered at $4.654$ $\mu m$ is the hydrogen line Pf $\beta$. The spectrum has been modeled by calculating the excitation of the rotational levels in the ground vibrational state with hydrogen as the collision partner (that is, not assuming an LTE solution; see Section \ref{sec:excitation}). The resultant absorption spectrum is plotted over the data (red dotted line).}
	\label{fig:BetaPic1}
\end{figure}

If hydrogen is the dominant collision partner, we can determine the density of \ion{H}{1} necessary to collisionally excite the observed rotational levels of CO. We adopt the distribution of \ion{H}{1} suggested by \citep{2004A&A...413..681B} and the collision rates between hydrogen and CO from \citet{1976ApJ...205..766G} using Equation \ref{eq:2}. We assume that the CO originates in an annulus embedded in a much larger radial distribution that includes hydrogen and other metals observed by \citep{2004A&A...413..681B}. That is, the CO is coming off of grains into a bath of diffuse gas, including hydrogen. We do not assume the CO and hydrogen have the same inner radius, but instead determine the density of hydrogen at the inner radius of CO (25 AU; from Section \ref{constraint}). Of course, the CO may originate at a slightly larger radius, and it isn't clear how wide the annulus is. The best-fitting density for the hydrogen at 25 AU in order to reproduce the CO populations observed is n$_{\rm H}=(2.5^{+7.1}_{-1.2})\times10^{5}$ cm$^{-3}$ (Figure \ref{fig:chisq_h}). Interestingly, the upper limit of the hydrogen density at the 3$\sigma$ level is unconstrained. Thus, assuming the levels are not in LTE sufficiently can describe the observations, possibly resolving the factor of $\sim$3 difference between the CO column density determined by \citet{2000ApJ...538..904R} and the column density presented in this work (Section \ref{sec:ro_vib_abs}). The fit to the absorption features is shown in Figure \ref{fig:BetaPic1}. There is a prediction for the R(3) feature, which seems to fit the 1.5$\sigma$ feature at the same location. If the feature is determined to be real with further observations, it would more strongly constrain our results, but it would not appreciably alter our conclusions. That is, the population would still prove to be sub-thermalized.

\subsection{Constraint on Disk Column Density and Mass}
Continuing with this scenario, we assume the CO is embedded in a much larger gas disk (Figure \ref{fig:diagram}) made up of hydrogen and other metals, and that the density of hydrogen we determined is the fiducial density at the mid-plane (z$=$0), and the hydrogen continues inward beyond the CO ($\ge$25 AU) to where the atomic gas originates (13 AU). If the hydrogen follows the density profile described using the distribution given by \citet[][Equation \ref{eq:distribution}]{2004A&A...413..681B} for other metals in the disk, then the fiducial \ion{H}{1} density scaled to 117 AU becomes n$_0$=(3.9$^{+11}_{-1.8}$)$\times$10$^4$ cm$^{-3}$. Note that the hydrogen density calculated in Section \ref{sec:excitation} is independent of the radial distribution of CO, but rather is calculated by fitting to the level populations. We can calculate the line-of-sight column density of atomic hydrogen by integrating through the disk and compare to observational limits. The line-of-sight hydrogen column density becomes:
\begin{eqnarray}
	N(H)&=&\int_{13}^{323} n_0 \left[\left(\frac{r}{117{\rm \;AU}}\right)^{2.4}+\left(\frac{r}{117{\rm \;AU}}\right)^{5.3}\right]^{-1/2}  dr\nonumber \\
		&=& (2.1^{+5.9}_{-1.0})\times10^{20} {\rm \;cm}^{-2}.
\end{eqnarray}

Integrating the distribution over the disk from 13 AU to 323 AU, we find that the total gas mass is: 
\begin{eqnarray}
	{\rm H~atoms}	&=&	2\pi \; \int_{13}^{323}\int_{-\infty}^\infty n_H(r,z)  r\;dr	dz	\nonumber\\
	 			&=&	(6.2^{+17}_{-2.9}) \times10^{50} {\rm~atoms} \nonumber\\
				&=&	(0.17^{+0.47}_{-0.08}) {\rm~M}_\oplus.
\end{eqnarray}
The mass is within the limits from 21 cm observations \citep[1.6 M$_\earth$;][]{1995A&A...301..231F}, but also within limits from dynamical arguments \citep[0.4 M$_\earth$][]{2005A&A...437..141T}. Thus we find that \ion{H}{1} mass required to excite the CO that we measure in absorption is close to the \ion{H}{1} mass inferred in the solar composition case from \citet{2004A&A...413..681B}.

\section{DISCUSSION} \label{sec:discussion}
We observe CO absorption that is sub-thermally populated in the disk around $\beta$~Pic. The observed properties of the absorption can be accounted for in two scenarios: one where H$_2$O and CO are being photodesorbed off of icy grains, and an alternative scenario in which the CO that is coming off icy grains is excited by collisions with a residual gaseous disk of primordial hydrogen. The scenario involving H$_2$O (or possibly another molecule) would result in a narrow region where the grains are being desorbed ($<$2.3 AU), while the scenario with residual, primordial hydrogen would result in a much larger region where the CO is being desorbed. 

The latter scenario requires that the CO and hydrogen are well mixed in the region where the CO is coming off the grains. The \ion{H}{1} mass required to excite the CO is just at the 0.1 M$_\earth$ required for ion-neutral collisions with hydrogen to explain the braking of the gas. Thus, it seems that while the excess carbon abundance is consistent with ion-ion collisions to be the breaking mechanism of the gas, the likely mass of the hydrogen in the disk is itself sufficient to brake the gas through ion-neutral collisions. The fact that the $\beta$ Pic disk is so long-lived is not surprising given the fact that there is enough mass to brake the gas. 

If hydrogen is the dominant collision partner, we speculate that the metals and CO observed in the disk around $\beta$ Pic are released from grains into a remaining reservoir of hydrogen gas. This may provide a possible explanation to the origin of the enhanced carbon abundance measured by \citet{2006Natur.441..724R}. Most of the metals (e.g. oxygen) may come from desorption of material throughout the disk as grains containing organic species with other metals desorb at the same rate as CO \citep{2009ApJ...693.1209O,2009A&A...496..281O}. However, if most of the oxygen is tied up in water while the carbon is in CH$_4$ and CO, then the lower sublimation temperatures of those species compared to water ($<100$ K versus 170 K for water) allow carbon-rich material to preferentially sublimate at the inner edge of the disk. Thus, carbon would be produced in excess relative to the other metals. 

The result presented here, especially the prediction for the mass of hydrogen, can be tested with more sensitive measurements of hydrogen. Deeper observations than \citet{1995A&A...301..231F} are necessary to test these predictions (an order of magnitude improvement would be sufficient to explore the estimated mass, while a factor of $\sim$50 improvement could rule out the lower limit to 3$\sigma$). The Square-Kilometer Aarray, estimated to be 50-100 times more sensitive than current radio telescopes \citep{2004NewAR..48..979C,2009IEEEP..97.1482D}, should be able to test these predictions.

\acknowledgments
Based in part on observations obtained at the Gemini Observatory, which is operated by the Association of Universities for Research in Astronomy, Inc., under a cooperative agreement with the NSF on behalf of the Gemini partnership: the National Science Foundation (United States), the Science and Technology Facilities Council (United Kingdom), the National Research Council (Canada), CONICYT (Chile), the Australian Research Council (Australia), MinistŽrio da Cincia e Tecnologia (Brazil) and SECYT (Argentina). The Phoenix infrared spectrograph was developed by the National Optical Astronomy Observatory. The Phoenix spectra were obtained as a part of programs GS-2002B-Q-43 and GS-2008A-C-7. Also based on observations obtained at the Infrared Telescope Facility, which is operated by the University of Hawaii under Cooperative Agreement no. NCC 5-538 with the National Aeronautics and Space Administration, Science Mission Directorate, Planetary Astronomy Program.
M.~R.~T. acknowledges this work was performed under contract with the Jet Propulsion Laboratory (JPL) funded by NASA through the Michelson Fellowship Program. JPL is managed for NASA by the California Institute of Technology. 

{\it Facilities:} \facility{Gemini:South (PHOENIX)}, \facility{IRTF (CSHELL)}.

\clearpage

\begin{deluxetable}{lccccc}
\tablecaption{Log of $\beta$ Pic Observations \label{tab:observations}}
\tablehead{
	\colhead{Date} & \colhead{Instrument} & \colhead{Spectral Range} & \colhead{Integration Time} & \colhead{S/N} & \colhead{Lines covered}\\
	 && (cm$^{-1}$) & (m)
}
\startdata
	2008 Mar 23 & PHOENIX & 2104 - 2113 & 16 & 90 & v=1-0: P(8), P(9), v=2-1: P(1), P(2), P(3)\\
	2003 Jan 12 & PHOENIX & 2150 - 2160 & 8 & 100 & v=1-0: R(1), R(2), R(3)\\
	2000 Aug 8 & CSHELL & 2146 - 2152 & 16 & 90 & v=1-0: R(0), R(1)
\enddata
\end{deluxetable}%

\begin{deluxetable}{lcccccc}
\tablecaption{Line Fluxes \label{tab:fluxes}}
\tablehead{
	\colhead{Line} & \colhead{$\tilde{\nu}_{rest}$} & \colhead{$\tilde{\nu}_{obs}$} & \colhead{V$_{rad}$} & \colhead{Equivalent Width} & N(CO) & \colhead{Flux} \\
	& (cm$^{-1}$) & (cm$^{-1}$) & (km s$^{-1}$) & ($\times10^{-3}$ cm$^{-1}$) & ($\times10^{14}$ cm$^{-2}$) & (ergs cm$^{-2}$ s$^{-1}$)
}
\startdata
	v=1-0 R(0) & 2147.08 & 2146.95 & 18 & $4.9\pm0.8$ & $4.3$ & \nodata \\
	v=1-0 R(1) & 2150.86 & 2150.71 & 21 & $6.0\pm0.5$ & $9.0$ & \nodata \\
	v=1-0 R(2) & 2154.59 & 2154.43 & 22 & $4.4\pm0.5$ & $7.2$ & \nodata \\
	v=1-0 R(3) & 2158.30 & 2158.14  & 22 & $<0.5\pm0.5$ & $<2.3$ & \nodata \\
	v=2-1 P(1) & 2112.98 & \nodata & \nodata & $<0.5$ & \nodata & $<7.1\times10^{-16}$	\\ 
	v=2-1 P(2) & 2109.14 & \nodata & \nodata & $<0.5$ & \nodata & $<7.1\times10^{-16}$  \\
	v=2-1 P(3) & 2105.26 & \nodata & \nodata & $<0.5$ & \nodata & $<7.1\times10^{-16}$
\enddata
\end{deluxetable}%

\clearpage

\end{document}